\documentclass[aps,prl,twocolumn,groupedaddress]{revtex4}

\usepackage{graphicx}
\usepackage{amsmath}
\usepackage{amssymb}
\usepackage{amsthm}
\usepackage{array}
\usepackage{xy}
\usepackage{subfigure}

\begin{document}


\title{Universal Jamming Phase Diagram in the Hard-Sphere Limit}


\author{Thomas K. Haxton}
\altaffiliation{Current address: Molecular Foundry, Lawrence Berkeley National Laboratory, Berkeley, CA 94720}
\author{Michael Schmiedeberg}
\altaffiliation{Current address: Heinrich-Heine-Universit\"at D\"usseldorf, 40225 D\"usseldorf, Germany}
\author{Andrea J. Liu}
\affiliation{Department of Physics and Astronomy, University of Pennsylvania, Philadelphia, PA, 19104}


\date{\today}

\begin{abstract}
We present a new formulation of the jamming phase diagram for a class of glass-forming fluids consisting of spheres interacting via finite-ranged repulsions at temperature $T$, packing fraction $\phi$ or pressure $p$, and applied shear stress $\Sigma$.  We argue that the natural choice of axes for the phase diagram are the dimensionless quantities $T/p\sigma^3$, $p\sigma^3/\epsilon$, and $\Sigma/p$, where $T$ is the temperature, $p$ is the pressure, $\Sigma$ is the stress, $\sigma$ is the sphere diameter, $\epsilon$ is the interaction energy scale, and $m$ is the sphere mass.  We demonstrate that the phase diagram is universal at low $p\sigma^3/\epsilon$; at low pressure, observables such as the relaxation time are insensitive to details of the interaction potential and collapse onto the values for hard spheres, provided the observables are non-dimensionalized by the pressure.  We determine the shape of the jamming surface in the jamming phase diagram, 
organize previous results in relation to the jamming phase diagram, and discuss the significance of various limits.
\end{abstract}


\maketitle

\section{I. Introduction}
Disordered solids made of molecules (glasses), microscale particles, droplets, or gas bubbles in fluid (colloidal glasses, emulsions, and foams), or solid macroscopic particles in air (granular materials) can be fluidized by raising temperature or kinetic energy, decreasing density or packing fraction, or  applying a mechanical load such as a shear stress.  These phenomena can be distilled into a ``jamming phase diagram" as a function of temperature, density, and shear stress~\cite{Liu1998}.  Such a diagram describes whether the system is ``jammed" ({\it i.~e.~} whether the relaxation time of the system exceeds some fixed, long time scale) or ``unjammed" ({\it i.~e.~} whether the relaxation time is shorter than that time scale).  While each system has a different jamming phase diagram, a jammed region generally exists at sufficiently low temperatures $T$, inverse packing fractions $\phi^{-1}$, or shear stresses $\Sigma$.

Recently, Xu et al.~\cite{Xu2009b} found that the equilibrium dynamical behavior of soft repulsive spheres is simplified considerably if one considers appropriate dimensionless quantities related to the relaxation time, temperature, pressure, and shear stress.  In particular, it is useful to express the relaxation time of the liquid $\tau$,  in terms of the time scale $\sqrt{m/p \sigma}$, where $m$ is the particle mass, $\sigma$ is the particle diameter, and $p$ is the pressure.  This time scale characterizes the time for a sphere to move a distance equal to its diameter when accelerated by a typical compressive force of $p \sigma^2$.  Thus, it provides an estimate of the duration of a particle rearrangement.  This choice is particularly convenient because if one defines the dynamic glass transition by the criterion $\tau \sqrt{p \sigma/m}=\textrm{constant}$, then in the low pressure limit, the glass transition is controlled solely by the ratio of temperature to pressure, $T/p \sigma^3$, and is independent of the potential.  Here, the low pressure limit corresponds to $p \sigma^3/\epsilon \ll 1$, where $\epsilon$ characterizes the repulsive interaction energy scale.  This limit also corresponds to the hard-sphere limit, where $\epsilon \rightarrow \infty$.  Thus, all soft repulsive spheres behave as hard spheres when $p \sigma^3/\epsilon \rightarrow 0$.

In this paper, we extend the analysis of Xu et al.~\cite{Xu2009b} to describe the behavior under steady-state shear.  The rheological behavior is characterized by the relation between the shear stress $\Sigma$, and the strain rate $\dot \gamma$.  Following the approach of Xu et al., we consider the dimensionless variables $\Sigma/p$ and $\dot \gamma \sqrt{m/p \sigma}$ and show that the steady-state shear rheology of soft repulsive spheres reduces to that of hard spheres in the limit where $p \sigma^3/\epsilon \rightarrow 0$.

In addition, we suggest a new formulation of the jamming phase diagram in terms of the dimensionless temperature, shear stress, and pressure, $T/p \sigma^3$, $\Sigma/p$, and $p \sigma^3/\epsilon$.  Here, the jamming surface is characterized by the criterion $\tau \sqrt{p \sigma/m}=\textrm{constant}$ instead of the standard criterion $\tau = \textrm{constant}$.  One advantage of this formulation is that the diagram is universal for soft repulsive spheres in the limit of low $p \sigma^3/\epsilon$.  
In addition, the jamming transition of packings of frictionless spheres, Point J~\cite{OHern2003}, which controls many of the properties of packings~\cite{Liu2010}, lies at the origin of the diagram.

\section{II.  Model and Methods} 
We consider a class of models of frictionless spheres with finite-range, repulsive interactions.  In particular, we consider bidisperse spheres of mass $m$, half with diameter $\sigma$ and half with diameter $1.4\sigma$.  The spheres interact through pairwise additive interaction potentials $V(r_{ij})$:
\begin{equation}
V_{\alpha}(r_{ij})=
\begin{cases} \dfrac{\epsilon}{\alpha}\left(1-\dfrac{r_{ij}}{\sigma_{ij}}\right)^\alpha & { \ \ \textrm{for}\ \ } r_{ij}<\sigma_{ij} \\
0 & { \ \ \textrm{for}\ \ } r_{ij} > \sigma_{ij},
\end{cases},
\label{pot}
\end{equation}
where $\sigma_{ij}=(\sigma_i+\sigma_j)/2$ is the separation at contact.  Note that $V_0(r)$ is the hard-sphere potential.

We study the steady-state shear rheology of three-dimensional systems by conducting non-equilibrium molecular dynamics simulations at fixed temperature $T$ and shear strain rate $\dot\gamma$.  Here, the Boltzmann constant $k_B$ is set to unity.  We define the temperature by the velocity fluctuations relative to an imposed uniform shear gradient and impose a fixed shear strain rate using Lees-Edwards boundary conditions; the system is sheared in the $x$-direction with the shear gradient in the $y$-direction.  We use periodic boundary conditions in the $z$-direction.  For hard spheres, we use an event-driven algorithm~\cite{Marin1993, Isobe1999} at fixed packing fraction $\phi$ and periodically rescale velocities to keep the temperature within $1\%$ of the desired value.  For soft spheres, we use a conventional molecular dynamics algorithm that numerically integrates classical equations of motion.  We employ Gaussian constraints~\cite{Evans1983a, Evans1983b, Hood1989} to fix the instantaneous temperature $T$ and pressure $p$.  In all cases, we measure the average steady-state shear stress $\Sigma$ and the relaxation time $\tau$ defined by $\Delta r_z(\tau)=\sigma/\sqrt{3}$, where $\Delta r_z(t) \equiv \sqrt{\langle (r_z(t)-r_z(0))^2\rangle}$ is the root-mean-squared displacement in the vorticity direction.

\section{III.  Rheology collapse in the low-pressure limit}
In this section we show that the rheology of repulsive soft spheres described by the potentials in Eq.~\ref{pot} reduces in the low-pressure limit to the rheology of hard spheres.  As discussed in the introduction, it is important to introduce the appropriate dimensionless quantities for describing the shear stress $\Sigma$ and the strain rate $\dot\gamma$.  The arguments of Xu et al.~\cite{Xu2009b}, suggest that it is most convenient to use the pressure to obtain the dimensionless quantities $\Sigma/p$ and $\dot\gamma \sqrt{m/p \sigma}$.   For a given potential, dimensional analysis tells us that we may write the dependence of the shear stress on the three control parameters, $T$, $p$, and $\dot\gamma$, as a dimensionless function $f$ of three independent dimensionless control parameters: 
\begin{equation}
\dfrac{\Sigma}{p}=f\left(\dfrac{T}{p \sigma^3}, \dot\gamma\sqrt{\dfrac{m}{p \sigma}}, \dfrac{p \sigma^3}{\epsilon}\right).
\label{rheoscale}
\end{equation}

Note that Eq.~\ref{rheoscale} isolates the interaction energy scale $\epsilon$ in only one of the three control parameters, the dimensionless pressure $p \sigma^3/\epsilon$.  The combinations $\Sigma/p$ and $\dot \gamma \sqrt{m/p \sigma}$ are familiar to the granular materials community~\cite{Forterre2008, daCruz2005, Peyneau2008}.  The dimensionless shear stress $\Sigma/p$ is a macroscopic dynamic friction coefficient, while the dimensionless strain rate $\dot \gamma \sqrt{m/p \sigma^3}$ is typically called the inertial number and is understood physically as follows.  As the system is sheared at constant pressure, it repeatedly dilates and contracts.  The dimensionless strain rate describes how fast the system is sheared relative to the time it takes for pressure to drive a dilated configuration into a close-packed configuration; it is the ratio of the contraction time to the shear time.

In the limit $p \sigma^3/\epsilon \rightarrow 0$, we expect
\begin{equation}
\dfrac{\Sigma}{p}=F\left(\dfrac{T}{p \sigma^3}, \dot\gamma\sqrt{\dfrac{m}{p \sigma}}\right).
\label{rheolimit}
\end{equation}

\begin{figure}
\includegraphics[width=8.5cm]{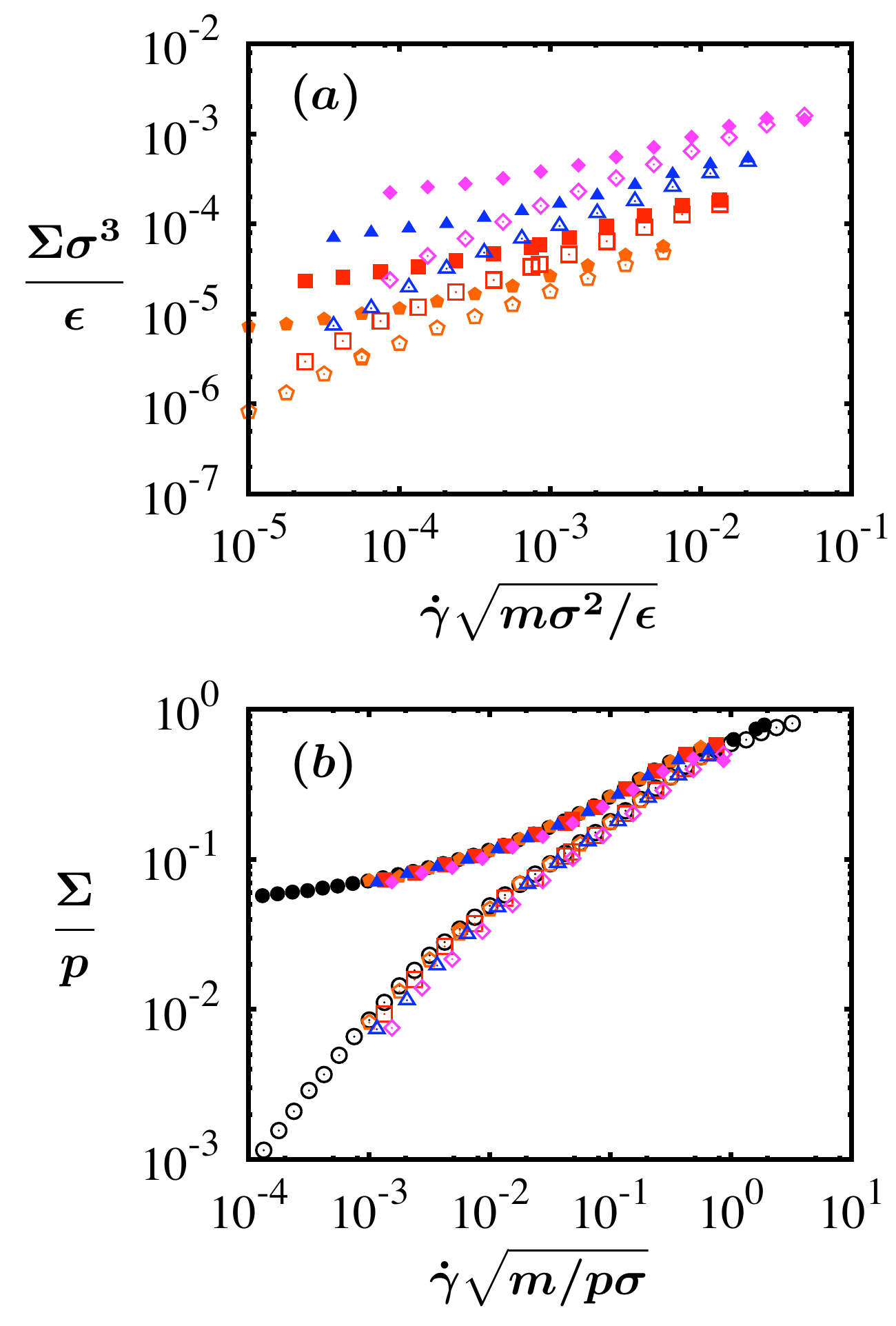}
\caption{(a) Shear stress, $\Sigma$, vs strain rate, $\dot\gamma$, in standard molecular dynamics units for $\alpha=2$, two different values of $T/p\sigma^3$, and four different values of $p\sigma^3/\epsilon$.  The filled symbols represent $T/p\sigma^3=0.03$ and the open symbols represent $T/p\sigma^3=0.1$.  The different shapes and colors represent different pressures: pink diamonds are $p\sigma^3/\epsilon=10^{-2.5}$; blue triangles are $p\sigma^3/\epsilon=10^{-3}$; red squares are $p\sigma^3/\epsilon=10^{-3.5}$; and orange pentagons are $p\sigma^3/\epsilon=10^{-4}$.  (b) Same data, made dimensionless by the pressure.  For comparison, data for hard spheres are also shown in black at $T/p\sigma^3=0.03$ (black filled circles), and at $T/p\sigma^3=0.1$ (black open circles).}
\label{fig:collapse}
\end{figure}

In Fig.~\ref{fig:collapse}, we demonstrate that the rheology for spheres with harmonic ($\alpha=2$) interactions collapses in the low-pressure limit onto the rheology for hard spheres, as expected.  In Fig.~\ref{fig:collapse} (a), we show the shear stress in standard simulation units, shear stress $\Sigma \sigma^3/\epsilon$ as a function of strain rate $\dot \gamma \sqrt{m \sigma^2/\epsilon}$.   Data are presented for spheres with harmonic repulsions ($\alpha=2$ in Eq.~\ref{pot}) at four different pressures, $p \sigma^3/\epsilon=10^{-4}, 10^{-3.5}, 10^{-3}$, and $10^{-2.5}$, and two different temperatures, $T/p \sigma^3=0.03$ and $T/p \sigma^3=0.1$.  In Fig.~\ref{fig:collapse} (b), we show that for each value of $T/p \sigma^3$ we can collapse the data from the four different pressures onto the hard sphere results by dividing the shear stress by the pressure and multiplying the strain rate by the time scale $\sqrt{m/p \sigma}$.  All the data at the lower dimensionless temperature, $T/p \sigma^3=0.03$, collapse onto a hard sphere curve with an apparent dynamic yield stress at low strain rates, while all the data at the higher dimensionless temperature, $T/p \sigma^3=0.1$, collapse onto a hard sphere curve with a linear viscous response, $\Sigma/p \propto \dot\gamma \sqrt{m/p \sigma}$, at low strain rates.  At higher strain rates, the system shear thins: $\Sigma/p$ grows more slowly than linearly with increasing $\dot \gamma \sqrt{m/p \sigma}$.  Note that for $0.03<T/p \sigma^3 < 0.1$ there is a continuum of curves, so the rheology changes continuously and is not described by two distinct branches, at least when plotted in this way.

\begin{figure}
\includegraphics[width=8.5cm]{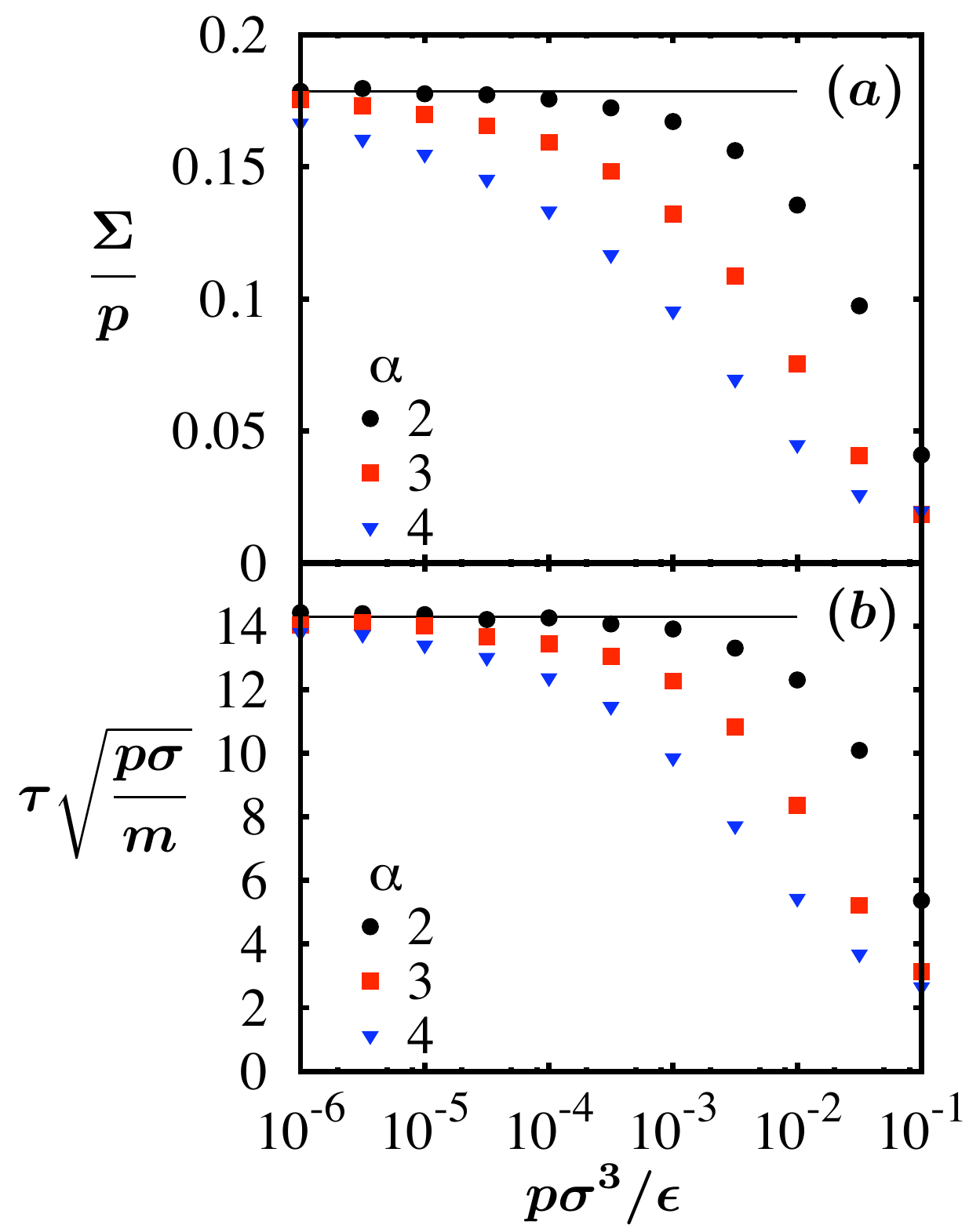}
\caption{Approach to the hard sphere limit.  (a) Dimensionless shear stress, $\Sigma/p$, vs pressure, $p$, at fixed dimensionless temperature, $T/p=0.1$, and dimensionless strain rate, $\dot\gamma/\sqrt{p}=0.1$, for three different exponents, $\alpha$.  (b) Dimensionless relaxation time, $\tau p^{1/2}$, vs pressure, $p$, for the same parameters, $T/p=0.1$, $\dot\gamma/\sqrt{p}=0.1$, and three different values of $\alpha$.  In each plot, the horizontal line represents the value for hard spheres.}
\label{fig:approach}
\end{figure}

As shown in Fig.~\ref{fig:approach} (a), the rheology approaches a well-defined limit as $p \sigma^3/\epsilon \rightarrow 0$ for several different potentials.  At fixed $T/p \sigma^3$ and $\dot\gamma \sqrt{m/p\sigma}$, $\Sigma/p$ approaches a limiting value as $p \sigma^3/\epsilon \rightarrow 0$.  While dimensional analysis does not require that this limit should be the same for different potentials, Fig.~\ref{fig:approach} (a) shows that $\Sigma/p$ approaches the hard-sphere value for several different exponents $\alpha$ (see Eq.~\ref{pot}) as $p \sigma^3/\epsilon \rightarrow 0$.  Thus, the function $F$ in Eq.~\ref{rheolimit} characterizes the rheology of hard spheres.  All potentials that vanish at a well-defined distance, such as the potentials of Eq.~\ref{pot}, reduce to the hard-sphere potential in the limit of zero overlap, so systems described by such potentials should exhibit hard-sphere behavior as $p \sigma^3/\epsilon \rightarrow 0$.  Eq.~\ref{rheolimit} therefore represents a universal limit of the rheology.

Note that the data collapse of Fig.~\ref{fig:collapse}(b) is not limited to the rheology but should apply to any observable quantity made dimensionless by the pressure.  For instance, as we show in fig.~\ref{fig:approach} (b), the dimensionless relaxation time $\tau \sqrt{p \sigma/m}$ also approaches its hard-sphere value as $p \sigma^3/\epsilon \rightarrow 0$.   

\section{IV. Nonlinear shear rheology of hard spheres near the glass transition}

\begin{figure}
\includegraphics[width=8.5cm]{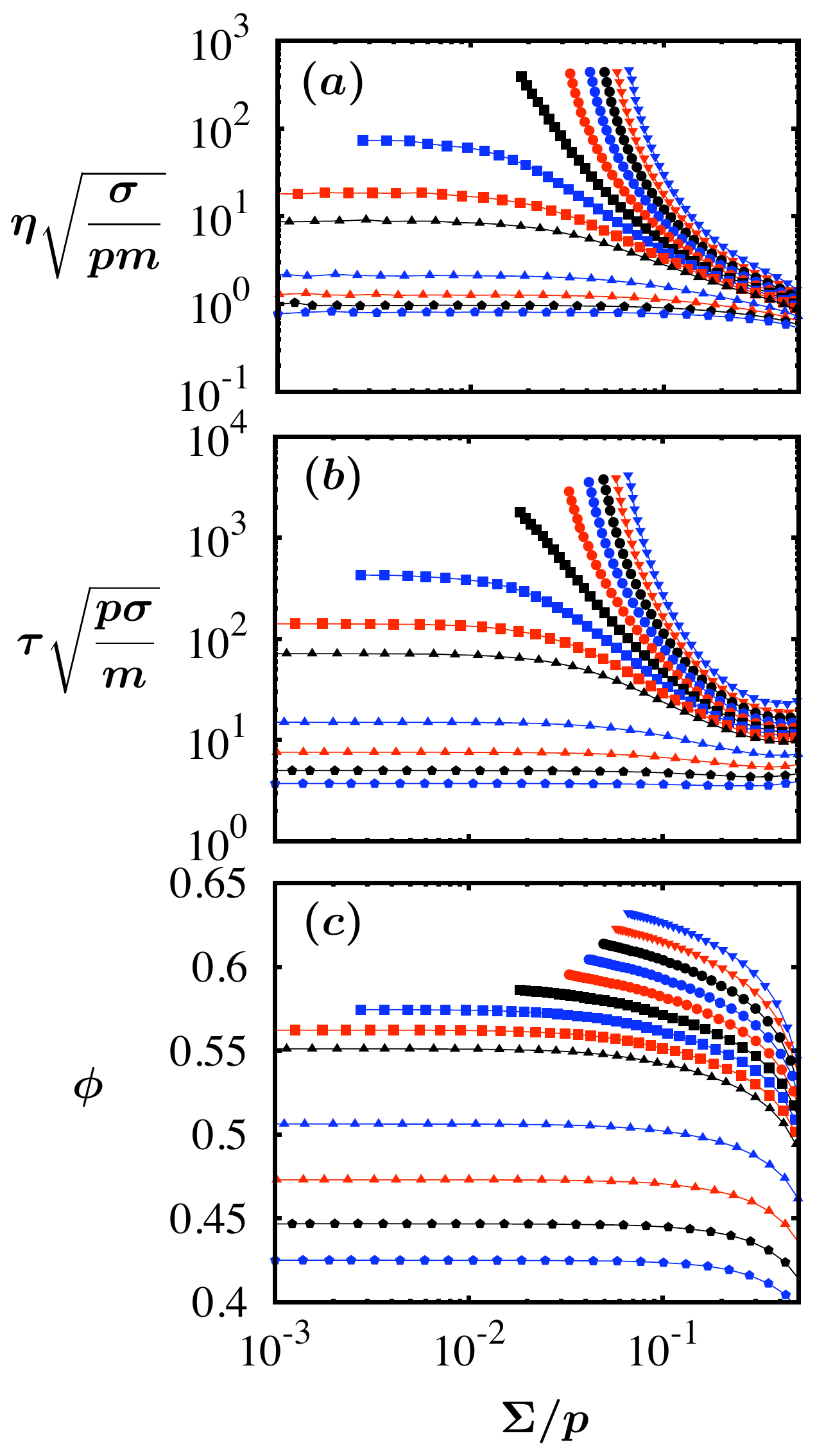}%
\caption{Hard sphere results interpolated at prescribed values of the dimensionless temperature, $T/p$.  (a) Dimensionless relaxation time, $\tau \sqrt{p \sigma/m}$, vs dimensionless shear stress, $\Sigma/p$. (b) Dimensionless shear viscosity, $\eta \sqrt{\sigma/p m}$, vs $\Sigma/p$. (c) Packing fraction, $\phi$, vs $\Sigma/p$.  Each connected line is a different value of $T/p\sigma^3$: 0.02 (blue down triangles), 0.03 (red down triangles), 0.04 (black circles), 0.05 (blue circles), 0.06 (red circles), 0.07 (black squares), 0.08 (blue squares), 0.09 (red squares), 0.1 (black up triangles), 0.15 (blue up triangles), 0.2 (red up triangles), 0.25 (black pentagons), and 0.3 (blue pentagons).}
\label{fig:hard}
\end{figure}

The results for soft spheres at low pressures show that it is important to understand the rheology and relaxation time of hard spheres in order to understand the corresponding behavior of a soft-sphere system.  We have performed non-equilibrium molecular dynamics simulations for hard spheres under shear, with results that are summarized in Fig.~\ref{fig:hard}.
In these plots, connected lines represent fixed values of $T/p \sigma^3$.  The results at fixed $T/p \sigma^3$ shown in Fig.~\ref{fig:hard} were obtained by interpolating from simulations conducted at fixed packing fraction.  In Fig.~\ref{fig:hard}(a), we plot the dimensionless shear viscosity $\eta \sqrt{\sigma/p m}$, defined as the ratio of the dimensionless shear stress $\Sigma/p$ to the dimensionless strain rate $\dot\gamma \sqrt{m/p \sigma}$, as a function of $\Sigma/p$.  In Fig.~\ref{fig:hard}(b), we plot the relaxation time $\tau \sqrt{p \sigma/m}$ vs~$\Sigma/p$.  The two plots are very similar: at high temperatures, both the viscosity and the relaxation time approach limiting values at low shear stress, while for low temperatures, the viscosity and relaxation time increase without bound on the time scales accessible to our simulations.  

Finally, Fig.~\ref{fig:hard}(c) shows the behavior of the packing fraction.  For a given value of $\Sigma/p$, $\phi$ decreases with increasing $T/p \sigma^3$, as expected; the kinetic energy pushes particles further apart at higher temperatures.  At high temperatures, $\phi$ is nearly independent of stress up to relatively high stresses, but at lower $T/p \sigma^3$, $\phi$ becomes more and more sensitive to $\Sigma/p$ because the system dilates ($\phi$ decreases) with increasing stress.

We have cut off the data in Fig.~\ref{fig:hard} at a dimensionless shear stress of $\Sigma/p \approx 0.5$ since there is a dramatic change in behavior there: $\tau \sqrt{p \sigma/m}$ increases while $\eta \sqrt{\sigma/pm}$, $\eta/p \tau$, and $\phi$ decrease sharply.  These changes are associated with the onset of layering of the spheres in the plane perpendicular to the shear gradient, as indicated by long-range order in the pair distribution function (not shown).  This layering facilitates the shearing of layers relative to each other but impedes the mobility of spheres within a layer, causing the shear viscosity to decouple from the relaxation time.  Such layering has been demonstrated to be an artifact of thermostats like ours that assume a linear shear profile~\cite{Evans1986, Delhommelle2003}.  Thermostats that do not assume a linear profile yield similar results below the layering transition but do not form layers at high stresses; instead, they exhibit shear thickening, where viscosity increases with shear stress.  At the layering transition, our system shear thickens at fixed packing fraction but not at fixed $T/p$.  Since the layering is an artifact of the thermostat, we focus on the isotropic phase below $\Sigma/p=0.5$.

\begin{figure}
\includegraphics[width=8.5cm]{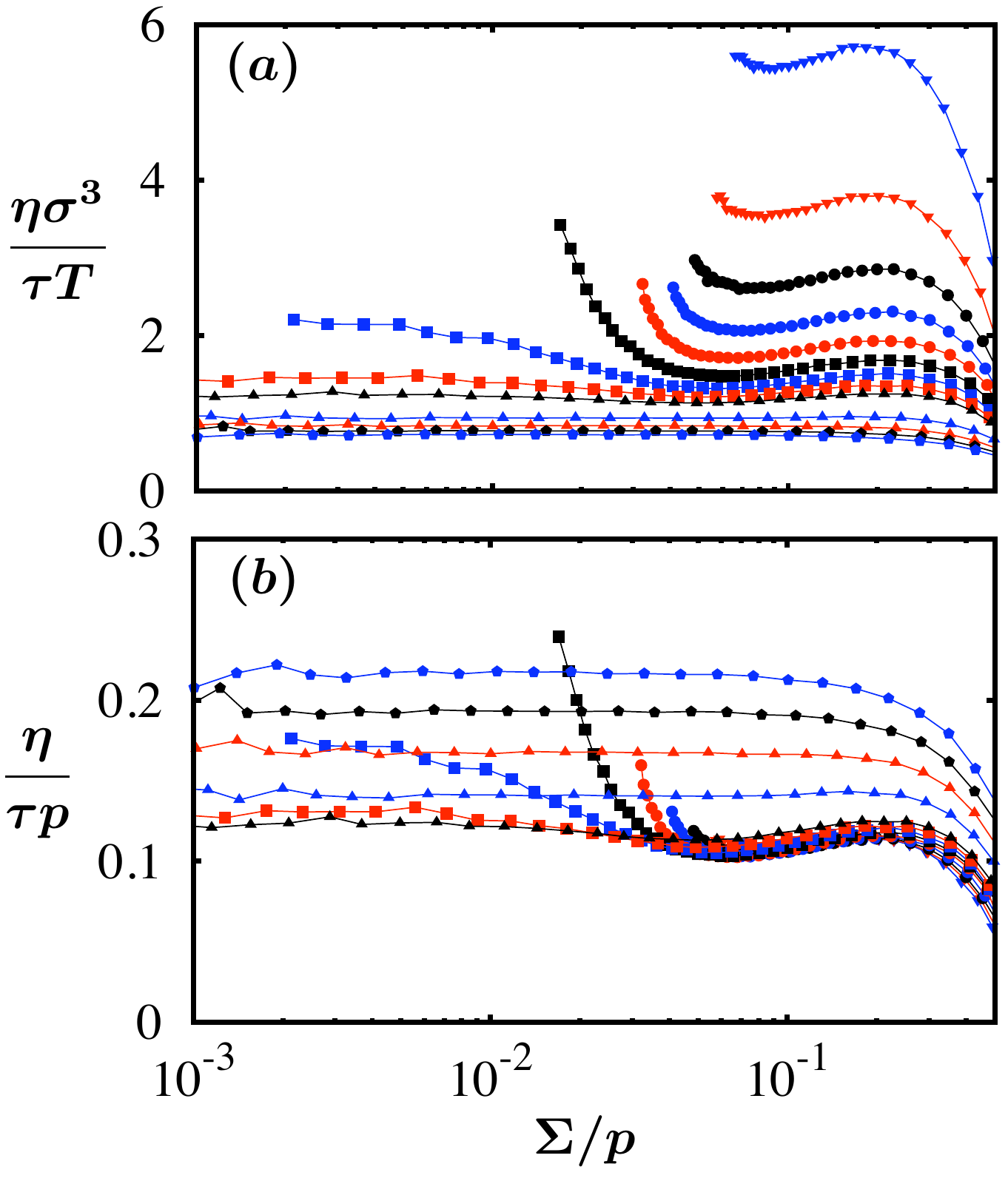}%
\caption{Ratio of shear viscosity to relaxation time for hard spheres.  As in Fig.~\ref{fig:hard}, each connected line represents a different value of $T/p\sigma^3$. (a) Ratio of shear viscosity to relaxation time in standard hard-sphere units, $\eta \sigma^3/T \tau$ vs $\Sigma/p$.  (b) Ratio of dimensionless shear viscosity, $\eta \sqrt{\sigma/pm}$, to dimensionless relaxation time, $\tau \sqrt{p \sigma/m}$.  As in Fig.~\ref{fig:hard}, the values of $T/p\sigma^3$ are: 0.02 (blue down triangles), 0.03 (red down triangles), 0.04 (black circles), 0.05 (blue circles), 0.06 (red circles), 0.07 (black squares), 0.08 (blue squares), 0.09 (red squares), 0.1 (black up triangles), 0.15 (blue up triangles), 0.2 (red up triangles), 0.25 (black pentagons), and 0.3 (blue pentagons).}
\label{fig:etatau}
\end{figure}

The viscosity and relaxation time are often used interchangeably to locate dynamic glass transitions.  This seems reasonable since Figs.~\ref{fig:hard}(a-b) show that their behavior is qualitatively similar.   In Fig.~\ref{fig:etatau}(a) we show the ratio of viscosity to relaxation time in standard hard-sphere units, $\eta \sigma^3/T \tau$.  Although the ratio is nearly independent of stress, the value of the ratio varies significantly with $T/p \sigma^3$.   Fig.~\ref{fig:etatau}(b) shows the ratio of $\eta \sqrt{\sigma/pm}$ to $\tau \sqrt{p \sigma/m}$, $\eta/\tau p$, in pressure units.  The result depends only weakly on $T/p \sigma^3$ over the range studied.  Thus, while viscosity can be used as a proxy for relaxation time for repulsive spheres at a given fixed pressure, care must be taken in comparing $\eta$ to $\tau$ if the pressure is varying.   

The near-collapse shown in Fig.~\ref{fig:etatau}(b) implies that up to relatively high shear stresses, the hard sphere fluid behaves like a Maxwell fluid with a relaxation time that decreases with shear stress but a modulus $\eta/\tau$ that is independent of shear stress and proportional to the pressure.  

\section{V. Dimensionless formulation of jamming phase diagram}

The simplicity of the behavior of repulsive spheres as a function of the dimensionless variables $T/p \sigma^3$, $\Sigma/p$, and $p \sigma^3/\epsilon$ suggests that it would be useful to recast the jamming phase diagram in terms of these variables instead of $T$, $\Sigma$, and packing fraction $\phi$~\cite{Liu1998}.  Both sets of variables span the same space of variables, but the choice  $\{T/p \sigma^3, \Sigma/p, p \sigma^3/\epsilon \}$ has a distinct advantage.  The plane at $p \sigma^3/\epsilon=0$, spanned by $T/p \sigma^3$ and $\Sigma/p$, defines the hard-sphere limit, which is universal for finite-ranged repulsions.  This choice of axes highlights the universality at $p \sigma^3/\epsilon=0$.  Another advantage of this choice of axes is that the zero-temperature jamming transition, Point J~\cite{OHern2003}, lies at the origin.

\begin{figure*}
\begin{center}
\includegraphics[width=17cm]{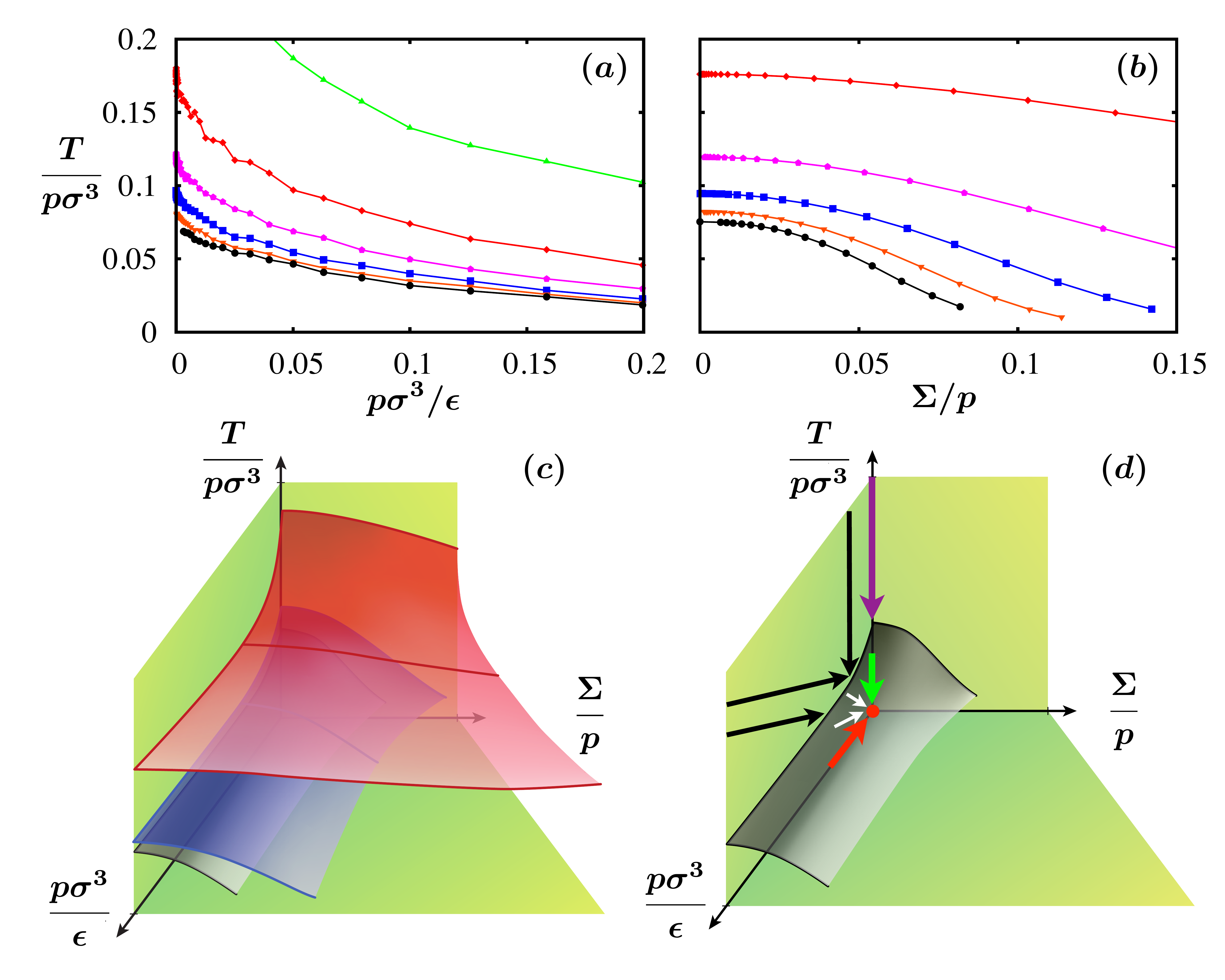}%
\end{center}
\caption{Jamming phase diagram for $\alpha=2$.  (a) Jamming phase diagram in the equilibrium plane at $\Sigma/p=0$ spanned by $\left\{ T/p \sigma^3, p \sigma^3/\epsilon \right\}$.  (b) Jamming phase diagram in the universal plane at $p\sigma^3/\epsilon=0$ spanned by $\left\{T/p \sigma^3, \Sigma/p \right\}$.  Panel (b) was constructed using hard spheres, which we showed to be equivalent to soft spheres at $p\sigma^3/\epsilon=0$.  In (a) and (b), we show contours of equal dimensionless relaxation time spaced by half decades: $\tau \sqrt{p \sigma/m}=10^{0.5}$ (yellow up triangles), $\tau \sqrt{p \sigma/m}=10$ (red diamonds), $\tau \sqrt{p \sigma/m}=10^{1.5}$ (pink pentagons), $\tau \sqrt{p \sigma/m}=10^{2}$ (blue squares), $\tau \sqrt{p \sigma/m}=10^{2.5}$ (orange down triangles), and $\tau \sqrt{p \sigma/m}=10^{3}$ (black circles).  The contours are constructed by interpolation.  (c) Full three-dimensional jamming phase diagram spanned by $\left\{ T/p \sigma^3, \Sigma/p, p \sigma^3/\epsilon \right\}$ for $\alpha=2$.  Lines represent logarithmically spaced contours of equal dimensionless relaxation time (top to bottom): $\tau \sqrt{p \sigma/m}=10$ (red), $\tau \sqrt{p \sigma/m}=10^{2}$ (blue), and $\tau \sqrt{p \sigma/m}=10^{3}$ (black).  We show contour lines along four planes cut through the diagram: the equilibrium plane at $\Sigma/p=0$, the hard-sphere plane at $p=0$, a plane at $p=0.1$, and a plane at $p=0.2$.  The surfaces are guides to the eye.  (d) Schematic illustration of various paths in the jamming phase diagram.  See text for details.
}
\label{fig:jpd}
\end{figure*}

It is instructive to first consider two-dimensional planes of the jamming phase diagram.  Fig.~\ref{fig:jpd} (a) shows contours of equal dimensionless relaxation time $\tau \sqrt{p \sigma/m}$, separated by half decades, in the equilibrium plane at $\Sigma/p=0$ spanned by $\left\{ T/p \sigma^3, p \sigma^3/\epsilon \right\}$ for a soft sphere system with a potential described by Eq.~\ref{pot} with $\alpha=2$.  The specific shape of these contours depends on the potential.  However, the downwards slope of the contours of  equal $\tau \sqrt{p \sigma/m}$ in Fig.~\ref{fig:jpd}(a) is more generic; it reflects the fact that at fixed $T/p \sigma^3$, the dimensionless relaxation time decreases with increasing $p \sigma^3/\epsilon$.  As the potential softens or the pressure increases at fixed $T/p \sigma^3$, the amount of overlap increases.  As a result, the soft spheres behave as hard spheres with a smaller diameter and relax more rapidly~\cite{Schmiedeberg2010}.  

Notice in Fig.~\ref{fig:jpd}(a) that the contours become more closely spaced as $T/p \sigma^3$ decreases.  This reflects the fact that spheres at fixed $p \sigma^3/\epsilon$ are fragile glass-formers; that is, $\log(\tau \sqrt{p \sigma/m})$ increases faster than linearly with $p \sigma^3/T$.  While the $T/p \sigma^3$-values of the contours decrease with increasing $p \sigma^3/\epsilon$, the relative spacing between contours remains similar, indicating that the fragility, defined in terms of the dimensionless variables $\tau \sqrt{p \sigma/m}$, $T/p \sigma^3$, and $p \sigma^3/\epsilon$ does not change significantly with increasing $p \sigma^3/\epsilon$.  If we instead define the fragility as the shape of the $\log(\tau \sqrt{p \sigma/m})$ vs $1/\phi$ curve, we find that the system becomes less fragile--that is, $\log(\tau \sqrt{p \sigma/m})$ vs $1/\phi$ increases less steeply--as the softness, $p \sigma^3/\epsilon$, increases, consistent with the interpretation of results for colloids of varied softness~\cite{Mattson2009}.

Fig.~\ref{fig:jpd}(b) shows contours of equal $\tau \sqrt{p \sigma/m}$, separated by half decades, in the plane at $p\sigma^3/\epsilon=0$ spanned by $\left\{T/p \sigma^3, \Sigma/p \right\}$.   As we noted earlier, the limit $p \sigma^3/\epsilon \rightarrow 0$ corresponds to the hard-sphere limit for any of the finite-ranged repulsions studied here.  Therefore, the contours of equal $\tau \sqrt{p \sigma/m}$ are universal in this plane.  We constructed Fig.~\ref{fig:jpd} (b) from hard-sphere data.  

Fig.~\ref{fig:jpd}(c) shows the full three-dimensional jamming phase diagram in $\{T/p \sigma^3, \Sigma/p, p \sigma^3/\epsilon\}$ space for the same soft-sphere system with $\alpha=2$ in Eq.~\ref{pot}.  We restrict ourselves to values of $T/p \sigma^3<0.2$, $\Sigma/p<0.15$, and $p \sigma^3/\epsilon<0.2$ to avoid the artifactual layering due to shear discussed in Sec.~IV.  Since we must interpolate to find the level sets of dimensionless relaxation time, it would be computationally expensive to construct entire surfaces.  Instead, we draw contours where the surfaces intersect four planes: the equilibrium plane at $\Sigma/p=0$, the hard sphere plane at $p \sigma^3/\epsilon$, a plane at $p \sigma^3/\epsilon=0.1$, and a plane at $p \sigma^3/\epsilon=0.2$.

The diagram in Fig.~\ref{fig:jpd}(c) differs from the standard jamming phase diagram in two ways.  First, the axes are different ($\{T/p \sigma^3, \Sigma/p, p \sigma^3/\epsilon\}$ instead of $\{ T, \Sigma, \phi \}$).  Second, the surfaces represent contours of equal {\it dimensionless} relaxation time, $\tau \sqrt{p \sigma/m}$, instead of contours of equal $\tau$ or $\tau \sqrt{\epsilon/m \sigma^2}$.  Here we have shown contours for three values of $\tau \sqrt{p \sigma/m}$, separated by decades.   The jamming surface for our system is given by the smallest surface closest to the origin and corresponds to $\tau \sqrt{p\sigma/m}=10^{-3}$ ; this is where we can no longer reach equilibrium on the time scale of our simulations.

As expected, the dimensionless relaxation time increases monotonically as any combination of the three control variables $\{T/p \sigma^3, \Sigma/p, p \sigma^3/\epsilon\}$ is reduced.  The contours become more closely spaced as $T/p \sigma^3$, $\Sigma/p$, or $p \sigma^3/\epsilon$ decrease, signaling the rapid increase of relaxation time as the system approaches jamming.  

The jamming surface intersects the unstressed plane ($\Sigma/p=0$) along a curve defined by $(T/p \sigma^3)_g=f_g(p \sigma^3/\epsilon)$, where $(T/p \sigma^3)_g$ is the dimensionless dynamic glass transition temperature.  This temperature depends on pressure, and decreases with increasing $p \sigma^3/\epsilon$.  However, recent results indicate that the relaxation time of soft spheres can be mapped onto the relaxation time for hard spheres by approximating the soft sphere potential by a hard sphere potential with a smaller effective diameter~\cite{Schmiedeberg2010}.  In particular, any approach to the jammed surface, such as the black arrows of Fig.~\ref{fig:jpd}(d), can be mapped onto the approach for hard spheres, the purple arrow of Fig.~\ref{fig:jpd}(d) along $p \sigma^3/\epsilon=\Sigma/p=0$.  This implies that the physics of the hard-sphere glass transition governs the glass transition of all finite-range repulsive spheres.

At nonzero shear stress, the jamming surface defines what could be considered either a stress-dependent dynamic glass transition temperature or a temperature-dependent dynamic yield stress $\Sigma_y$.  For the hard-sphere system at $p \sigma^3/\epsilon=0$, there is a nonzero dynamic yield stress for $T/p \sigma^3 < (T/p \sigma^3)_g^{HS}$, where 
$(T/p \sigma^3)_g^{HS}=f_g(p\sigma^3/\epsilon=0)$ 
marks the dynamic glass transition for unstressed hard spheres.  At all pressures, the contours of constant relaxation time $\tau \sqrt{p \sigma/m}$ must be quadratic in $\Sigma/p$ at small $\Sigma/p$.  This is expected from symmetry; the magnitude of the glass transition temperature should not depend on the sign of the shear stress.  Thus, if the shape of the dynamic yield stress curve for hard spheres is given by $\Sigma_y/p = f_{HS}((T/p \sigma^3)_g^{HS} - (T/p \sigma^3))$, then $f(x) \sim \sqrt{x}$ for small $x$.  A similar shape has been observed for contours of equal viscosity for a model metallic glass-forming liquid~\cite{Guan2010}.

Now consider the low temperature limit, $T/p \sigma^3 \rightarrow 0$.  In this limit, the shear stress $\Sigma/p$ appears to saturate at a nonzero value set by the zero-temperature dynamic yield stress $\Sigma_{y0}$ whose value depends on pressure: $\Sigma_{y0}/p=f_0(p\sigma^3/\epsilon)$, where $f_0(x)>0$.   In the hard-sphere limit, where $p \sigma^3/\epsilon \rightarrow 0$, the low-temperature system has a nonzero dynamic yield stress (Fig.~\ref{fig:jpd}(d)): $f_0(p\sigma^3/\epsilon)=f_0(p\sigma^3/\epsilon=0)={\rm constant}$.   Thus, at low values of $p \sigma^3/\epsilon$ for soft spheres, the athermal dynamic yield stress $\Sigma_{y0}$ must scale with pressure:
$\Sigma_{y0} \sim p$ at $T/p \sigma^3=0$.  This is consistent with granular experiments and simulations that find a macroscopic dynamic friction coefficient in the limit of low strain rate~\cite{Forterre2008, daCruz2005, Peyneau2008, Hatano2008, Hatano2010}.    

We may use the jamming phase diagram to characterize other dimensionless observables besides $\tau \sqrt{p \sigma/m}$, such as the packing fraction $\phi$, as functions of the dimensionless control parameters $\{T/p \sigma^3, \Sigma/p, p \sigma^3/\epsilon\}$.   For example, in the fluid region each value of $\{T/p \sigma^3, \Sigma/p, p \sigma^3/\epsilon\}$ corresponds to a certain value of the packing fraction; this is the equation of state.  We find that $\phi$ increases approximately linearly with $p^{1/(\alpha-1)}$, in agreement with the scaling for static sphere packings at $T/p \sigma^3=\Sigma/p=0$~\cite{OHern2003}.  The packing fraction decreases approximately linearly with $T/p \sigma^3$, consistent with the idea that $T/p \sigma^3$ controls the amount of free volume.  It also decreases nearly linearly with $\Sigma/p$, but with a smaller coefficient.

In the jammed region below the jamming surface, many studies have focused on the behavior at $T/p=\Sigma/p=0$ as packing fraction is reduced toward Point J, the point at $T=\Sigma=0$ and $\phi=\phi_c\approx 0.64$ where most static sphere packings lose mechanical stability~\cite{Liu2010}.  This path corresponds to decreasing $p \sigma^3/\epsilon$ toward the origin along $T/p \sigma^3=\Sigma/p=0$, as shown by the red arrow in Fig.~\ref{fig:jpd}(d).   Brito and Wyart~\cite{Brito2006, Brito2007, Brito2009} have argued that a hard-sphere glass ({\it i.e.} a system at $p \sigma^3/\epsilon=0$ below the gray surface in Fig.~\ref{fig:jpd}(d)) can effectively be mapped onto a jammed, athermal soft-sphere one at time scales short compared to the time between rearrangements due to aging.  The effective coordination number and vibrational properties of the system along the path described by the green arrow at $p \sigma^3/\epsilon=\Sigma/p=0$ can be mapped onto those along the red arrow at $T/p \sigma^3=\Sigma/p=0$~\cite{Brito2006, Brito2007, Brito2009}.   It is likely that the short-time-scale properties of repulsive spheres along any path that approaches the origin of the jamming phase diagram, such as the paths described by the white arrows in Fig.~\ref{fig:jpd}(d),  are controlled by the same physics. 

However, in the jammed region the value of the packing fraction at a given state point $\{T/p \sigma^3, \Sigma/p, p \sigma^3/\epsilon\}$ is not given by an equation of state but becomes history dependent.  For example, if $T/p \sigma^3$ is decreased slowly at $\Sigma/p=p\sigma^3/\epsilon=0$, then the resulting value of $\phi$ may be higher than if $T/p \sigma^3$ were decreased quickly to the same final value.  The equation of state becomes multi-valued; at least one extra parameter, for instance the packing fraction, must be specified in order to determine the state of the system.  Within the mean-field theory of random first-order models, glass states are uniquely determined by the jamming state diagram obtained by adding one more parameter to the jamming phase diagram~\cite{Mari2009}, but in general the state of the glass may depend in more detail on the history.

Note that because the jammed state is history-dependent, the value of $\phi_c$ that corresponds to the jamming transition at the origin of the jamming phase diagram, $\{T/p \sigma^3=0, \Sigma/p=0, p \sigma^3/\epsilon=0\}$, is not unique.  However, the approach to the origin in $\{T/p \sigma^3, \Sigma/p, p \sigma^3/\epsilon\}$ space is well defined.  This is another advantage of this formulation of the jamming phase diagram.

Finally, we note that there is one region of parameter space that cannot be explored in the reformulated version.  All systems at strictly zero temperature, zero applied stress, and packing fractions below the jamming transition, $\phi<\phi_c$, have zero pressure, so these states cannot be uniquely determined by the values of $\{T/p \sigma^3, \Sigma/p, p \sigma^3/\epsilon\}$.  However, as long as the temperature or stress are not strictly zero, the pressure becomes nonzero, and the system can be represented on the reformulated diagram.

\section{VI. Summary}

We have demonstrated that the jamming transition that occurs as a function of some combination of temperature, pressure or packing fraction, and applied mechanical load is conveniently described in terms of the dimensionless quantities $T/p\sigma^3$, $p \sigma^3/\epsilon$, and $\Sigma/p$.  Such a formulation defines the three-dimensional state space spanned by $T$, $p$, and $\Sigma$ as a product of the hard-sphere plane at $p \sigma^3/\epsilon=0$ and the unstressed plane at $\Sigma/p=0$.  One advantage of this formulation is that all repulsive spheres act as hard spheres near $p \sigma^3/\epsilon=0$, so the jamming surfaces for different repulsive spheres collapse on that plane.  A second advantage is that any repulsive sphere system will undergo a dynamic jamming transition as any combination of $\{T/p \sigma^3, \Sigma/p, p \sigma^3/\epsilon\}$ is decreased before reaching Point J at the origin.  

While the jamming surface is universal at low $p \sigma^3/\epsilon$, it depends on the shape of the interaction potential at higher $p \sigma^3/\epsilon$.  At zero shear stress, it has recently been shown that the relaxational dynamics of repulsive spheres can be mapped onto the dynamics of hard spheres using only structural information~\cite{Schmiedeberg2010}, even for high values of $p \sigma^3/\epsilon$.  These results suggest that, at least for repulsive spheres, the entire $\{T/p \sigma^3, p \sigma^3/\epsilon\}$ plane at $\Sigma/p=0$ can be mapped onto the hard-sphere path denoted by the purple arrow in Fig.~\ref{fig:jpd}.  Previous work suggests that a reasonably well-defined effective temperature controls the behavior of soft spheres under shear stress~\cite{Berthier2002, Ono2002, OHern2004, Ilg2007, Haxton2007}.   One open question is whether the relaxation time of a system at nonzero $\Sigma/p$, corresponding to some value of $T_{\rm eff}/p \sigma^3$, can be mapped onto the equilibrium hard-sphere behavior at the corresponding value of $T/p \sigma^3$. 

Finally, we note that the results presented here apply to repulsive spheres only.   In particular, the result that the jamming phase diagram is universal at low $p \sigma^3/\epsilon$ does not hold for systems with attractions.  In order to describe jamming of molecular liquids, we must also understand the effect of attractive interactions on the jamming phase diagram. 

\begin{acknowledgments}
We thank B. G. Chen, T. Egami, S. R. Nagel, and N. Xu for useful discussions.  This work was funded by DOE DE-FG02-05ER46199 (AJL and TKH), the UPENN-MRSEC DMR-0520020 (TKH), and the German Academic Exchange Service (DAAD) within the postdoc program (MS) .
\end{acknowledgments}

\bibliographystyle{apsrev.bst}

\begin{thebibliography}{28}
\expandafter\ifx\csname natexlab\endcsname\relax\def\natexlab#1{#1}\fi
\expandafter\ifx\csname bibnamefont\endcsname\relax
  \def\bibnamefont#1{#1}\fi
\expandafter\ifx\csname bibfnamefont\endcsname\relax
  \def\bibfnamefont#1{#1}\fi
\expandafter\ifx\csname citenamefont\endcsname\relax
  \def\citenamefont#1{#1}\fi
\expandafter\ifx\csname url\endcsname\relax
  \def\url#1{\texttt{#1}}\fi
\expandafter\ifx\csname urlprefix\endcsname\relax\def\urlprefix{URL }\fi
\providecommand{\bibinfo}[2]{#2}
\providecommand{\eprint}[2][]{\url{#2}}

\bibitem[{\citenamefont{Liu and Nagel}(1998)}]{Liu1998}
\bibinfo{author}{\bibfnamefont{A.~J.} \bibnamefont{Liu}} \bibnamefont{and}
  \bibinfo{author}{\bibfnamefont{S.~R.} \bibnamefont{Nagel}},
  \bibinfo{journal}{Nature (London)} \textbf{\bibinfo{volume}{396}},
  \bibinfo{pages}{21} (\bibinfo{year}{1998}).

\bibitem[{\citenamefont{Xu et~al.}(2009)\citenamefont{Xu, Haxton, Liu, and
  Nagel}}]{Xu2009b}
\bibinfo{author}{\bibfnamefont{N.}~\bibnamefont{Xu}},
  \bibinfo{author}{\bibfnamefont{T.~K.} \bibnamefont{Haxton}},
  \bibinfo{author}{\bibfnamefont{A.~J.} \bibnamefont{Liu}}, \bibnamefont{and}
  \bibinfo{author}{\bibfnamefont{S.~R.} \bibnamefont{Nagel}},
  \bibinfo{journal}{Phys. Rev. Lett.} \textbf{\bibinfo{volume}{103}},
  \bibinfo{pages}{245701} (\bibinfo{year}{2009}).

\bibitem[{\citenamefont{O'Hern et~al.}(2003)\citenamefont{O'Hern, Silbert, Liu,
  and Nagel}}]{OHern2003}
\bibinfo{author}{\bibfnamefont{C.~S.} \bibnamefont{O'Hern}},
  \bibinfo{author}{\bibfnamefont{L.~E.} \bibnamefont{Silbert}},
  \bibinfo{author}{\bibfnamefont{A.~J.} \bibnamefont{Liu}}, \bibnamefont{and}
  \bibinfo{author}{\bibfnamefont{S.~R.} \bibnamefont{Nagel}},
  \bibinfo{journal}{Phys. Rev. E} \textbf{\bibinfo{volume}{68}},
  \bibinfo{pages}{011306} (\bibinfo{year}{2003}).

\bibitem[{\citenamefont{Liu and Nagel}(2010)}]{Liu2010}
\bibinfo{author}{\bibfnamefont{A.~J.} \bibnamefont{Liu}} \bibnamefont{and}
  \bibinfo{author}{\bibfnamefont{S.~R.} \bibnamefont{Nagel}},
  \bibinfo{journal}{Annu. Rev. Condens. Matt. Phys.}
  \textbf{\bibinfo{volume}{1}} (\bibinfo{year}{2010}).

\bibitem[{\citenamefont{Mar\/in et~al.}(1993)\citenamefont{Mar\/in, Risso, and
  Cordero}}]{Marin1993}
\bibinfo{author}{\bibfnamefont{M.}~\bibnamefont{Mar\/in}},
  \bibinfo{author}{\bibfnamefont{D.}~\bibnamefont{Risso}}, \bibnamefont{and}
  \bibinfo{author}{\bibfnamefont{P.}~\bibnamefont{Cordero}},
  \bibinfo{journal}{J. Comp. Phys.} \textbf{\bibinfo{volume}{109}},
  \bibinfo{pages}{306} (\bibinfo{year}{1993}).

\bibitem[{\citenamefont{Isobe}(1999)}]{Isobe1999}
\bibinfo{author}{\bibfnamefont{M.}~\bibnamefont{Isobe}}, \bibinfo{journal}{Int.
  J. Mod. Phys. C} \textbf{\bibinfo{volume}{10}}, \bibinfo{pages}{1281}
  (\bibinfo{year}{1999}).

\bibitem[{\citenamefont{Evans and Morriss}(1983{\natexlab{a}})}]{Evans1983a}
\bibinfo{author}{\bibfnamefont{D.~J.} \bibnamefont{Evans}} \bibnamefont{and}
  \bibinfo{author}{\bibfnamefont{G.~P.} \bibnamefont{Morriss}},
  \bibinfo{journal}{Chem. Phys.} \textbf{\bibinfo{volume}{77}},
  \bibinfo{pages}{63} (\bibinfo{year}{1983}{\natexlab{a}}).

\bibitem[{\citenamefont{Evans and Morriss}(1983{\natexlab{b}})}]{Evans1983b}
\bibinfo{author}{\bibfnamefont{D.~J.} \bibnamefont{Evans}} \bibnamefont{and}
  \bibinfo{author}{\bibfnamefont{G.~P.} \bibnamefont{Morriss}},
  \bibinfo{journal}{Phys. Lett.} \textbf{\bibinfo{volume}{98A}},
  \bibinfo{pages}{433} (\bibinfo{year}{1983}{\natexlab{b}}).

\bibitem[{\citenamefont{Hood et~al.}(1989)\citenamefont{Hood, Evans, and
  Hanley}}]{Hood1989}
\bibinfo{author}{\bibfnamefont{L.~M.} \bibnamefont{Hood}},
  \bibinfo{author}{\bibfnamefont{D.~J.} \bibnamefont{Evans}}, \bibnamefont{and}
  \bibinfo{author}{\bibfnamefont{H.~J.~M.} \bibnamefont{Hanley}},
  \bibinfo{journal}{J. Stat. Phys.} \textbf{\bibinfo{volume}{57}},
  \bibinfo{pages}{729} (\bibinfo{year}{1989}).

\bibitem[{\citenamefont{Forterre and Pouliquen}(2008)}]{Forterre2008}
\bibinfo{author}{\bibfnamefont{Y.}~\bibnamefont{Forterre}} \bibnamefont{and}
  \bibinfo{author}{\bibfnamefont{O.}~\bibnamefont{Pouliquen}},
  \bibinfo{journal}{Annu. Rev. Fluid Mech.} \textbf{\bibinfo{volume}{40}},
  \bibinfo{pages}{1} (\bibinfo{year}{2008}).

\bibitem[{\citenamefont{da~Cruz et~al.}(2005)\citenamefont{da~Cruz, Emam,
  Prochnow, Roux, and Chevoir}}]{daCruz2005}
\bibinfo{author}{\bibfnamefont{F.}~\bibnamefont{da~Cruz}},
  \bibinfo{author}{\bibfnamefont{S.}~\bibnamefont{Emam}},
  \bibinfo{author}{\bibfnamefont{M.}~\bibnamefont{Prochnow}},
  \bibinfo{author}{\bibfnamefont{J.-N.} \bibnamefont{Roux}}, \bibnamefont{and}
  \bibinfo{author}{\bibfnamefont{F.}~\bibnamefont{Chevoir}},
  \bibinfo{journal}{Phys. Rev. E} \textbf{\bibinfo{volume}{72}},
  \bibinfo{pages}{021309} (\bibinfo{year}{2005}).

\bibitem[{\citenamefont{Peyneau and Roux}(2008)}]{Peyneau2008}
\bibinfo{author}{\bibfnamefont{P.-E.}~\bibnamefont{Peyneau}} \bibnamefont{and}
  \bibinfo{author}{\bibfnamefont{J.-N.}~\bibnamefont{Roux}},
  \bibinfo{journal}{Phys. Rev. E} \textbf{\bibinfo{volume}{78}},
  \bibinfo{pages}{011307} (\bibinfo{year}{2008}).

\bibitem[{\citenamefont{Evans and Morriss}(1986)}]{Evans1986}
\bibinfo{author}{\bibfnamefont{D.~J.} \bibnamefont{Evans}} \bibnamefont{and}
  \bibinfo{author}{\bibfnamefont{G.~P.} \bibnamefont{Morriss}},
  \bibinfo{journal}{Phys. Rev. Lett.} \textbf{\bibinfo{volume}{56}},
  \bibinfo{pages}{2172} (\bibinfo{year}{1986}).

\bibitem[{\citenamefont{Delhommelle et~al.}(2003)\citenamefont{Delhommelle,
  Petravic, and Evans}}]{Delhommelle2003}
\bibinfo{author}{\bibfnamefont{J.}~\bibnamefont{Delhommelle}},
  \bibinfo{author}{\bibfnamefont{J.}~\bibnamefont{Petravic}}, \bibnamefont{and}
  \bibinfo{author}{\bibfnamefont{D.~J.} \bibnamefont{Evans}},
  \bibinfo{journal}{Phys. Rev. E} \textbf{\bibinfo{volume}{68}},
  \bibinfo{pages}{031201} (\bibinfo{year}{2003}).

\bibitem[{\citenamefont{Schmiedeberg et~al.}()\citenamefont{Schmiedeberg,
  Haxton, Nagel, and Liu}}]{Schmiedeberg2010}
\bibinfo{author}{\bibfnamefont{M.}~\bibnamefont{Schmiedeberg}},
  \bibinfo{author}{\bibfnamefont{T.~K.} \bibnamefont{Haxton}},
  \bibinfo{author}{\bibfnamefont{S.~R.} \bibnamefont{Nagel}}, \bibnamefont{and}
  \bibinfo{author}{\bibfnamefont{A.~J.} \bibnamefont{Liu}}, \bibinfo{note}{in
  preparation}.

\bibitem[{\citenamefont{Mattson et~al.}(2009)\citenamefont{Mattson, Wyss,
  Fernandez-Nieves, Miyazaki, Hu, Reichman, and Weitz}}]{Mattson2009}
\bibinfo{author}{\bibfnamefont{J.}~\bibnamefont{Mattson}},
  \bibinfo{author}{\bibfnamefont{H.~M.} \bibnamefont{Wyss}},
  \bibinfo{author}{\bibfnamefont{A.}~\bibnamefont{Fernandez-Nieves}},
  \bibinfo{author}{\bibfnamefont{K.}~\bibnamefont{Miyazaki}},
  \bibinfo{author}{\bibfnamefont{Z.}~\bibnamefont{Hu}},
  \bibinfo{author}{\bibfnamefont{D.~R.} \bibnamefont{Reichman}},
  \bibnamefont{and} \bibinfo{author}{\bibfnamefont{D.~A.} \bibnamefont{Weitz}},
  \bibinfo{journal}{Nature (London)} \textbf{\bibinfo{volume}{462}},
  \bibinfo{pages}{83} (\bibinfo{year}{2009}).

\bibitem[{\citenamefont{Guan et~al.}(2010)\citenamefont{Guan, Chen, and
  Egami}}]{Guan2010}
\bibinfo{author}{\bibfnamefont{P.}~\bibnamefont{Guan}},
  \bibinfo{author}{\bibfnamefont{M.}~\bibnamefont{Chen}}, \bibnamefont{and}
  \bibinfo{author}{\bibfnamefont{T.}~\bibnamefont{Egami}},
  \bibinfo{journal}{Phys. Rev. Lett.} \textbf{\bibinfo{volume}{104}},
  \bibinfo{pages}{205701} (\bibinfo{year}{2010}).

\bibitem[{\citenamefont{Hatano}(2008)}]{Hatano2008}
\bibinfo{author}{\bibfnamefont{T.}~\bibnamefont{Hatano}}, \bibinfo{journal}{J.
  Phys. Soc. Jpn.} \textbf{\bibinfo{volume}{77}}, \bibinfo{pages}{123002}
  (\bibinfo{year}{2008}).

\bibitem[{\citenamefont{Hatano}(2010)}]{Hatano2010}
\bibinfo{author}{\bibfnamefont{T.}~\bibnamefont{Hatano}},
  \bibinfo{journal}{Progr. Theoret. Phys. Suppl.}
  \textbf{\bibinfo{volume}{184}}, \bibinfo{pages}{143} (\bibinfo{year}{2010}).

\bibitem[{\citenamefont{Brito and Wyart}(2006)}]{Brito2006}
\bibinfo{author}{\bibfnamefont{C.}~\bibnamefont{Brito}} \bibnamefont{and}
  \bibinfo{author}{\bibfnamefont{M.}~\bibnamefont{Wyart}},
  \bibinfo{journal}{Europhys. Lett.} \textbf{\bibinfo{volume}{76}},
  \bibinfo{pages}{149} (\bibinfo{year}{2006}).

\bibitem[{\citenamefont{Brito and Wyart}(2007)}]{Brito2007}
\bibinfo{author}{\bibfnamefont{C.}~\bibnamefont{Brito}} \bibnamefont{and}
  \bibinfo{author}{\bibfnamefont{M.}~\bibnamefont{Wyart}}, \bibinfo{journal}{J.
  Stat. Mech: Theory and Experiment} p. \bibinfo{pages}{L08003}
  (\bibinfo{year}{2007}).

\bibitem[{\citenamefont{Brito and Wyart}(2009)}]{Brito2009}
\bibinfo{author}{\bibfnamefont{C.}~\bibnamefont{Brito}} \bibnamefont{and}
  \bibinfo{author}{\bibfnamefont{M.}~\bibnamefont{Wyart}}, \bibinfo{journal}{J.
  Chem. Phys.} \textbf{\bibinfo{volume}{131}}, \bibinfo{pages}{024504}
  (\bibinfo{year}{2009}).

\bibitem[{\citenamefont{Mari et~al.}(2009)\citenamefont{Mari, Krzakala, and
  Kurchan}}]{Mari2009}
\bibinfo{author}{\bibfnamefont{R.}~\bibnamefont{Mari}},
  \bibinfo{author}{\bibfnamefont{F.}~\bibnamefont{Krzakala}}, \bibnamefont{and}
  \bibinfo{author}{\bibfnamefont{J.}~\bibnamefont{Kurchan}},
  \bibinfo{journal}{Phys. Rev. Lett.} \textbf{\bibinfo{volume}{103}},
  \bibinfo{pages}{025701} (\bibinfo{year}{2009}).

\bibitem[{\citenamefont{Berthier and Barrat}(2002)}]{Berthier2002}
\bibinfo{author}{\bibfnamefont{L.}~\bibnamefont{Berthier}} \bibnamefont{and}
  \bibinfo{author}{\bibfnamefont{J.-L.} \bibnamefont{Barrat}},
  \bibinfo{journal}{J. Chem. Phys.} \textbf{\bibinfo{volume}{116}},
  \bibinfo{pages}{6228} (\bibinfo{year}{2002}).

\bibitem[{\citenamefont{Ono et~al.}(2002)\citenamefont{Ono, O'Hern, Durian,
  Langer, Liu, and Nagel}}]{Ono2002}
\bibinfo{author}{\bibfnamefont{I.~K.} \bibnamefont{Ono}},
  \bibinfo{author}{\bibfnamefont{C.~S.} \bibnamefont{O'Hern}},
  \bibinfo{author}{\bibfnamefont{D.~J.} \bibnamefont{Durian}},
  \bibinfo{author}{\bibfnamefont{S.~A.} \bibnamefont{Langer}},
  \bibinfo{author}{\bibfnamefont{A.~J.} \bibnamefont{Liu}}, \bibnamefont{and}
  \bibinfo{author}{\bibfnamefont{S.~R.} \bibnamefont{Nagel}},
  \bibinfo{journal}{Phys. Rev. Lett.} \textbf{\bibinfo{volume}{89}},
  \bibinfo{pages}{095703} (\bibinfo{year}{2002}).

\bibitem[{\citenamefont{O'Hern et~al.}(2004)\citenamefont{O'Hern, Liu, and
  Nagel}}]{OHern2004}
\bibinfo{author}{\bibfnamefont{C.~S.} \bibnamefont{O'Hern}},
  \bibinfo{author}{\bibfnamefont{A.~J.} \bibnamefont{Liu}}, \bibnamefont{and}
  \bibinfo{author}{\bibfnamefont{S.~R.} \bibnamefont{Nagel}},
  \bibinfo{journal}{Phys. Rev. Lett.} \textbf{\bibinfo{volume}{93}},
  \bibinfo{pages}{165702} (\bibinfo{year}{2004}).

\bibitem[{\citenamefont{Ilg and Barrat}(2007)}]{Ilg2007}
\bibinfo{author}{\bibfnamefont{P.}~\bibnamefont{Ilg}} \bibnamefont{and}
  \bibinfo{author}{\bibfnamefont{J.-L.} \bibnamefont{Barrat}},
  \bibinfo{journal}{Europhys. Lett.} \textbf{\bibinfo{volume}{79}},
  \bibinfo{pages}{26001} (\bibinfo{year}{2007}).

\bibitem[{\citenamefont{Haxton and Liu}(2007)}]{Haxton2007}
\bibinfo{author}{\bibfnamefont{T.~K.} \bibnamefont{Haxton}} \bibnamefont{and}
  \bibinfo{author}{\bibfnamefont{A.~J.} \bibnamefont{Liu}},
  \bibinfo{journal}{Phys. Rev. Lett.} \textbf{\bibinfo{volume}{99}},
  \bibinfo{pages}{195701} (\bibinfo{year}{2007}).

\end{thebibliography}

\end{document}